\documentclass[aps,twocolumn,prd,superscriptaddress,longbibliography,reprint]{revtex4-2}
\usepackage{amsfonts}
\usepackage{mathrsfs}
\usepackage{amsmath}
\usepackage{color}
\usepackage{graphicx}
\usepackage{bm}
\usepackage{amssymb}
\usepackage{xspace}
\usepackage{epstopdf}
\usepackage{longtable}
\usepackage{multirow}
\usepackage{booktabs}
\usepackage{textcomp}
\usepackage[title]{appendix}
\usepackage{float}
\usepackage{dsfont}
\usepackage[colorlinks=true, letterpaper=true, pdfstartview=FitV, linkcolor=blue, citecolor=blue, urlcolor=blue]{hyperref}
\usepackage[]{changes}
\definechangesauthor[name={parkman}, color=red]{XuTao}

\begin{document}

\title{Three-dimensional real Chern insulator in bulk $\gamma$-graphyne}

\author{Xu-Tao Zeng}
\affiliation{School of Physics, Beihang University, Beijing 100191, China}

\author{Bin-Bin Liu}
\affiliation{School of Physics, Beihang University, Beijing 100191, China}

\author{Fan Yang}
\affiliation{School of Physics, Beihang University, Beijing 100191, China}

\author{Zeying Zhang}
\affiliation{College of Mathematics and Physics, Beijing University of Chemical Technology, Beijing 100029, China}
\affiliation{Research Laboratory for Quantum Materials, Singapore University of Technology and Design, Singapore 487372, Singapore}



\author{Y. X. Zhao}
\affiliation{National Laboratory of Solid State Microstructures and Department of Physics, Nanjing University, Nanjing 210093, China}
\affiliation{Collaborative Innovation Center of Advanced Microstructures, Nanjing University, Nanjing 210093, China}

\author{Xian-Lei Sheng}
\email{xlsheng@buaa.edu.cn}
\affiliation{School of Physics, Beihang University, Beijing 100191, China}
\affiliation{Peng Huanwu Collaborative Center for Research and Education, Beihang University, Beijing 100191, China}

\author{Shengyuan A. Yang}
\affiliation{Research Laboratory for Quantum Materials, Singapore University of Technology and Design, Singapore 487372, Singapore}

\begin{abstract}
The real Chern insulator state, featuring nontrivial real Chern number and second-order boundary modes, has been revealed in a few two-dimensional systems. The concept can be extended to three dimensions (3D), but a proper material realization is still lacking. Here, based on first-principles calculations and theoretical analysis, we identify the recently synthesized bulk $\gamma$-graphyne as a 3D real Chern insulator. Its nontrivial bulk topology leads to
topological hinge modes spreading across the 1D edge Brillouin zone. Under compression of the interlayer distance,
the system can undergo a topological phase transition into a real nodal-line semimetal, which hosts three bulk nodal rings and topological boundary modes on both surfaces and hinges.
We also develop a minimal model which captures essential physics of the system.

\end{abstract}

\maketitle

\section{Introduction}

Graphynes refer to a family of van der Waals (vdW) layered carbon allotropes with $sp$ and $sp^2$ hybridization, first predicted by Baughman \emph{et al.} in 1987~\cite{baughman1987}. Like graphite, each layer of a graphyne structure is flat with one-atom-thickness, and it can be constructed from a layer of graphite (i.e., monolayer graphene) by replacing certain C-C bonds with acetylenic groups~\cite{diederich1994,gao2019,li2010}. In 2010, a member of the graphyne family, graphdiyne, was synthesized in experiment by Li \emph{et al.}~\cite{li2010}, which generated great research interest~\cite{hirsch2010,kang2019}. Meanwhile, $\gamma$-graphyne, which is the simplest and the representative member of the family, also attracted tremendous experimental efforts. For a long time, it was only realized in the form of small fragments without long range order~\cite{diederich1994,bunz1999}. In a recent breakthrough, Hu \emph{et al.}~\cite{hu2022} demonstrated the synthesis of \emph{bulk} crystalline $\gamma$-graphyne by a novel reversible dynamic alkyne metathesis approach and revealed its $ABC$ stacking order.

Previous works have proposed or demonstrated many excellent properties of these materials~\cite{jiao2011,ouyang2012,ajori2013,chen2013,tang2014,ren2015,zhao2015,li2015a,yang2013a,xue2018,solis2019,azizi2020}. Notably, many members of the graphyne family also possess nontrivial topological states~\cite{malko2012,ahn2018,sheng2019,lee2020,liu2019e,chen2021,chen2022,liu2022,zhu2022}. For example, several semimetallic monolayer graphynes were predicted to host Dirac points similar to graphene~\cite{malko2012}. More interestingly, although these carbon allotropes do not support the conventional topological insulator state due to their negligible spin-orbit coupling (SOC)~\cite{shen2012,bernevig2013}, the presence of inversion ($\mathcal{P}$) and time reversal ($\mathcal{T}$) symmetries permit a class of \emph{real} topological phases. Namely, the band eigenstates are essentially real valued, for which the topological classification differs from the complex ones~\cite{zhao2016,zhao2017}.
In light of this, monolayer graphdiyne and $\gamma$-graphyne were revealed as first examples of a two-dimensional (2D)
real Chern insulator (RCI), characterized by a nontrivial real Chern number $\nu_R=1$ and featured with topological corner zero-modes~\cite{sheng2019,chen2021}. Three-dimensional (3D) bulk graphdiyne was
predicted as the first example of second-order nodal-line semimetals~\cite{chen2022}, and its nodal lines exhibit a linking structure~\cite{ahn2018}. Recently, it was found that 2D RCI states also generally exist in the phonon spectra of monolayer graphyne family materials~\cite{zhu2022}.

A common way to construct a 3D topological state is via stacking 2D topological states~\cite{fu2007a,burkov2011a}. Then, what are possible consequences of stacking together 2D RCI layers? When the interlayer coupling is strong, there could appear a band inversion along the stacking direction, which results in a topological semimetal state~\cite{wang2020}. The gap closing manifold, which separates the $\nu_R=1$ and $\nu_R=0$ $k_z$-slices (assuming the stacking is along the $z$ direction), generally has the form of a nodal loop~\cite{wang2020}. This state corresponds to the second-order nodal-line semimetal found in 3D bulk graphdiyne~\cite{chen2022}.
On the other hand, when the interlayer coupling is weak, there will be no band inversion along $k_z$, and the system will become a 3D RCI~\cite{chen2022}. Such a state was predicted for phonons in bulk graphdiyne~\cite{zhu2022}. Regarding electronic systems, it was suggested also in bulk graphdiyne, but requiring an unrealistic tensile strain~\cite{chen2022}. Therefore, it remains a challenge to identify a real material that realizes the 3D RCI state.

Motivated by the recent success in synthesizing bulk $\gamma$-graphyne~\cite{hu2022} and by the quest for 3D RCI materials, in this work, we explore the electronic properties of bulk $\gamma$-graphyne and reveal it as the first example of an intrinsic electronic 3D RCI.
Using first-principles calculations, we evaluate real Chern numbers for the band structure and identify the 3D RCI state. The nontrivial topology in the bulk manifests as topological boundary modes located at a pair of hinges of a bulk sample. We show that under strain, bulk $\gamma$-graphyne can transition from 3D RCI into a topological semimetal phase with three nodal rings. However, different from bulk graphdiyne, the nodal rings here do not carry a nontrivial real Chern number, but they still feature a nontrivial 1D winding number, which leads to protected surface modes. Finally, we construct a minimal model that captures the essential physics of bulk $\gamma$-graphyne, which can serve as a starting point for further studies of this interesting material.

\section{Crystal structure}

\begin{figure}
  \includegraphics[width=8 cm]{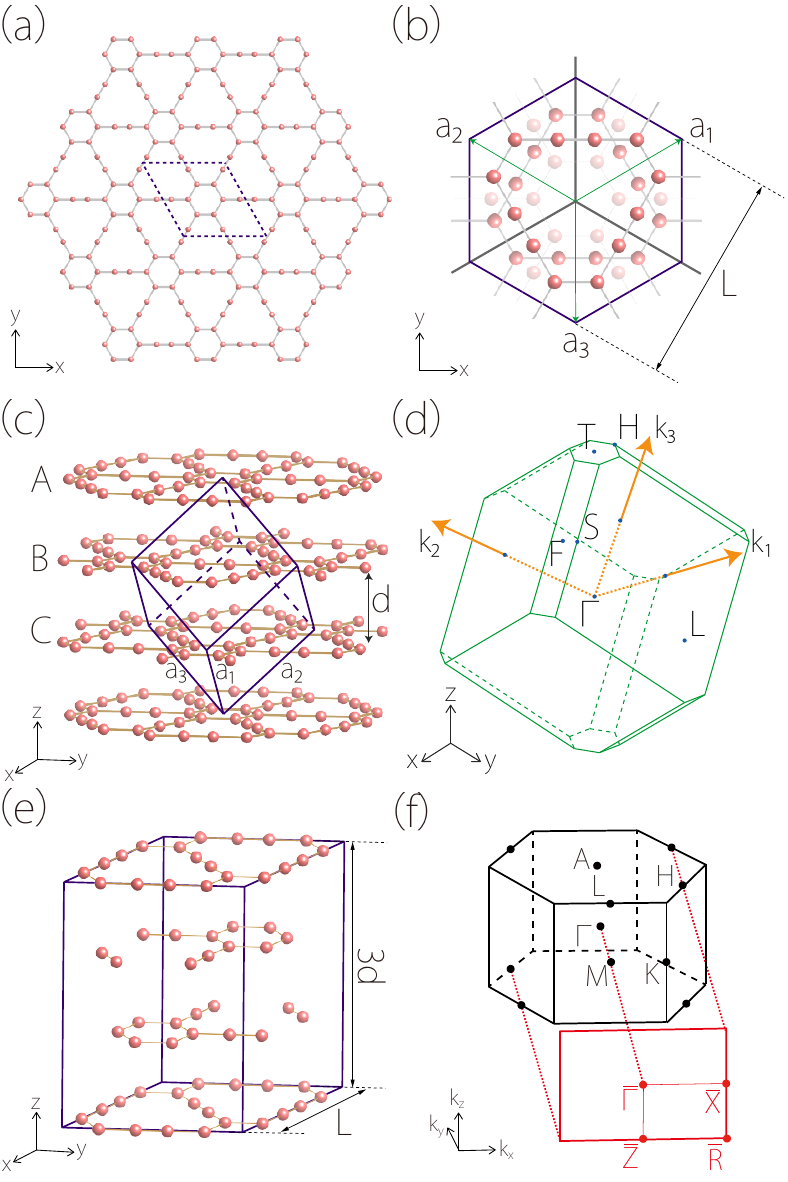}
  \caption{(a) Crystal structure of a single $\gamma$-graphyne layer. (b) Top view of the structure of 3D $\gamma$-graphyne. Three layers in a unit cell are indicated with different transparency. $L$ is the lattice parameter.  (c) Perspective view of bulk $\gamma$-graphyne. Three primitive lattice vectors are indicated by $a_{1,2,3}$. $d$ is the interlayer distance. (d) BZ for the primitive cell. (e) The conventional unit cell used for studying boundary modes.
  (f)  BZ for the conventional cell in (e). The surface BZ for a side surface (normal to $\hat{y}$) is also displayed. }
  \label{fig:structure}
\end{figure}

Figure~\ref{fig:structure}(a) illustrates the lattice structure of a single $\gamma$-graphyne layer extracted from the 3D bulk. It is completely flat and can be viewed as a 2D hexagonal network of benzene rings ($sp^2$ hybridized) connected by acetylenic (-C$\equiv$C-) linkages ($sp$ hybridized). Bulk $\gamma$-graphyne consists of such monolayers stacked along the out-of-plane direction (denoted as the $z$ direction). The recent experiment determined the $ABC$ stacking order in bulk $\gamma$-graphyne~\cite{hu2022}, as illustrated in Fig.~\ref{fig:structure}(b, c).
This 3D structure belongs to the trigonal crystal system, with space group No.~166 ($R \overline{3} m$). The $ABC$ stacking preserves the $C_{3z}$ symmetry, but breaks $C_{2z}$ (which exists in the monolayer). Hence, it reduces the point group symmetry from $D_{6h}$ to $D_{3d}$. Importantly, bulk $\gamma$-graphyne preserves both $\mathcal{P}$ and $\mathcal{T}$ symmetries, and as a carbon material, it has negligible SOC. It follows that for such a system, $\mathcal{PT}$ symmetry constrains the eigenstates (and the Hamiltonian) to be real. This fulfills the condition required for real topological phases.

In our first-principles calculations (details are presented in Appendix A), we use the experimental lattice parameters with $L=6.9$ \AA ~and $d=3.5$ \AA~\cite{hu2022} [see Fig.~\ref{fig:structure}(b, c)].

In calculating the bulk band structure (in Fig.~2), we adopt the primitive cell as shown in Fig.~\ref{fig:structure}(c), which has a rhombohedral shape and contains 12 carbon atoms. The corresponding Brillouin zone (BZ) is shown in Fig.~\ref{fig:structure}(d).
{In addition, when studying boundary modes, it is easier to use the conventional cell in Fig.~\ref{fig:structure}(e), whose volume is three times that of the primitive cell. The corresponding hexagonal prism shaped BZ is shown in Fig.~\ref{fig:structure}(f).}

\section{Band structure and 3D real Chern insulator}

\begin{figure}
  \includegraphics[width=8 cm]{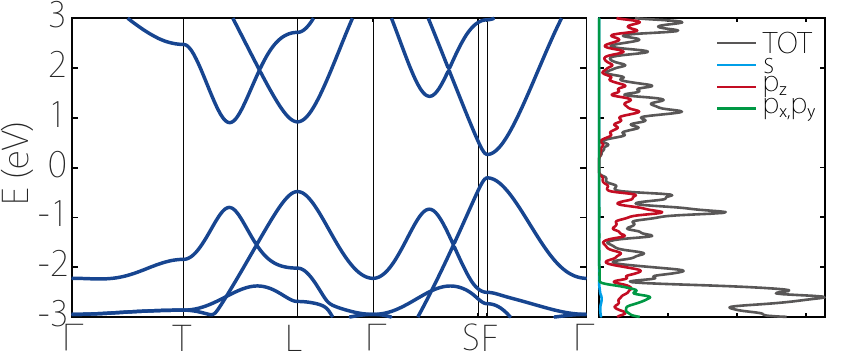}
  \caption{Calculated band structure and PDOS for bulk $\gamma$-graphyne (calcualted with primitive cell).}
  \label{fig:band}
\end{figure}

Our calculated band structure for bulk $\gamma$-graphyne is shown in Fig.~\ref{fig:band}, together with the projected density of states (PDOS). Since the system is a semiconductor, the result here is computed with the hybrid functional
(HSE06)~\cite{perdew1996} for better accuracy.  From Fig.~\ref{fig:band}(a), we see that  bulk $\gamma$-graphyne is a direct gap semiconductor with the gap located at the $F$ point (there are six $F$ points in BZ related by the $S_{6z}$ symmetry). From the calculated PDOS, one observes that the low-energy states of bulk $\gamma$-graphyne, i.e., the states around band edges, are dominated by the carbon $p_z$ orbitals. The gap value is $\sim 0.47$ eV at HSE06 level. We find that with HSE06+GW correction~\cite{onida2002}, the gap value can be further increased to $\sim 0.91$ eV (see Supplemental Material). Nevertheless, we have checked that the different gap values from different approaches
(also including commonly used functionals based on generalized gradient approximation) do not alter the band topology, so in the following, we will base our discussion on the HSE06 result.  As we mentioned, SOC is negligible in carbon materials. We have checked the band structure with SOC, which shows little difference, so SOC is dropped in the discussion below and the system can be regarded as spinless.

\begin{table}[thpb]
  \caption{Parity eigenvalues at the eight $\mathcal{P}$-invariant points. The coordinates of these points are given by $T$ (0.5, 0.5, 0.5), $L_1$ (0.5, 0, 0), $L_2$ (0, 0.5, 0), $L_3$ (0, 0, 0.5), $F_1$ (0.5, -0.5, 0), $F_2$ (0, 0.5, -0.5), and $F_3$ (0.5, 0, -0.5). $n_{+}$ ($n_{-}$) denotes the number of occupied bands with positive (negative) parity eigenvalues.}
  \begin{tabular}{ccccclcccc}
  \hline \hline
  \multirow{2}{*}{{Parity}}
  & \multicolumn{4}{c}{$k_z=0$}             &
  & \multicolumn{4}{c}{$k_z=\pi$} \\
  \cline{2-5} \cline{7-10}
  & $\Gamma$  & $F_1$    & $F_2$    & $F_3$ &
  & $T$       & $L_1$    & $L_2$    & $L_3$
  \\ \hline
  $n_{ +}$
  & 11 & 13 & 13 & 13                    &
  & 11   & 13  & 13   & 13   \\
  $n_{-}$
  & 13 & 11 & 11 & 11                    &
  & 13   & 11   & 11   & 11   \\
  \hline
  $\nu_R$
  & \multicolumn{4}{c}{$1$}             &
  & \multicolumn{4}{c}{$1$} \\
  \hline \hline
  \end{tabular}
  \label{tab:TRIM}
\end{table}


Since bulk $\gamma$-graphyne is effectively a  spinless system, it cannot be a conventional 3D topological insulator, which must require SOC. Nevertheless, due to the presence of $\mathcal{PT}$ symmetry, it may realize a real topological insulator.
To investigate the topological character, we evaluate the real Chern number $\nu_R$ for a 2D slice of BZ with fixed $k_z$.
Actually, since the system has a {global} band gap, all such 2D $k_z$-slices are adiabatically connected to each other and hence they must share the same $\nu_R$. This means we can evaluate $\nu_R$ by choosing a particular slice.

Here, we consider the two slices at $k_z=0$ and $k_z=\pi$. Either slice contains four $\mathcal{P}$-invariant momentum points $\Gamma_i$. Specifically, for $k_z=0$, the four points are $\Gamma$ and three $F$ points (labeled as $F_1$, $F_2$, and $F_3$); whereas for  $k_z=\pi$, they are given by $T$ and three $L$ points (labeled as $L_1$, $L_2$, and $L_3$).
For such a slice, $\nu_R$ can be readily extracted  from the parity eigenvalues at the four $\Gamma_i$ points, from the formula~\cite{ahn2018,chen2021}
\begin{equation}
  (-1)^{\nu_R}=\prod_{i} (-1)^{\left\lfloor (n^{\Gamma_i}_-/2)\right\rfloor},
\end{equation}
where $\lfloor\cdots\rfloor$ is the floor function and $n^{\Gamma_i}_-$ is the number of occupied bands with negative parity eigenvalue at $\Gamma_i$. The values of $n^{\Gamma_i}_-$ obtained from our first-principles calculations are listed in Table~\ref{tab:TRIM}. From the result, one finds that $k_z=0$ and $k_z=\pi$ planes have the same $\nu_R=1$, as it should be. This nontrivial real Chern number indicates that bulk $\gamma$-graphyne is indeed a 3D RCI.

%

\section{Surface spectrum and hinge modes}

In Refs.~\cite{wang2020,chen2021}, it has been shown that a 2D RCI features an interesting bulk-boundary correspondence. In the general case, a nontrivial $\nu_R$ in the 2D bulk indicates second-order topological boundary modes, namely, the 1D edges are gapped whereas gapless zero-modes exist at the 0D corners. Now, a 3D RCI, such as the bulk $\gamma$-graphyne, can be viewed as
formed by 2D RCIs stacked along the $z$ direction. Hence, intuitively, we would expect that 2D surfaces of 3D RCI should be gapped, whereas the topological zero-modes should appear at 1D hinges of a sample.

\begin{figure}[tb]
  \includegraphics[width=8 cm]{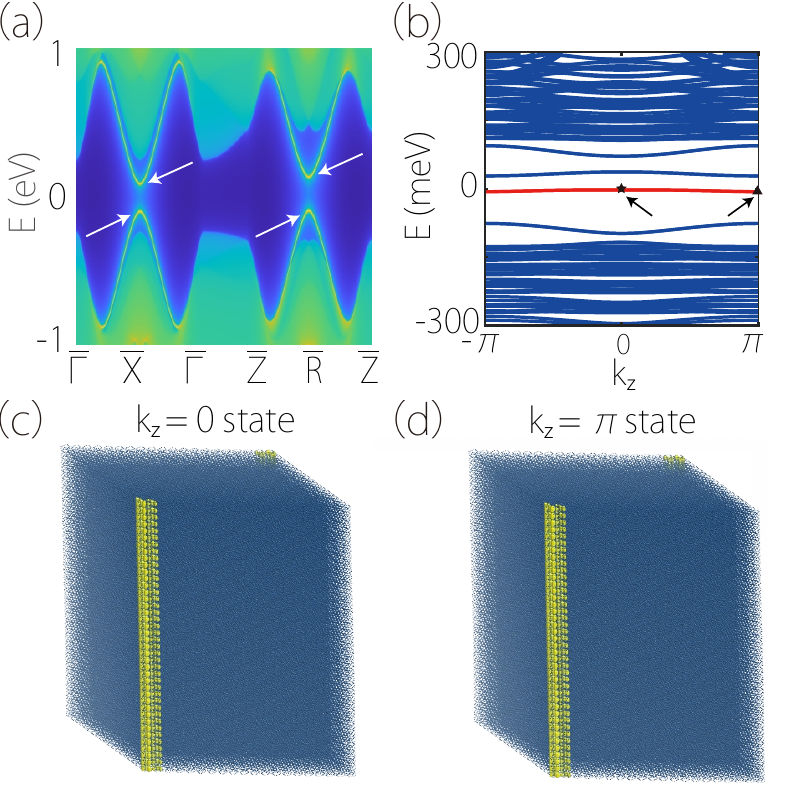}
  \caption{(a) Projected spectrum for the side surface normal to $\hat{y}$ (see Fig.~\ref{fig:structure}(f)). The plot is with respect to the conventional cell. (b) Spectrum for a sample with the tube-like geometry in (c). The red color indicates the hinge modes. The spatial distributions for the doubly degenerate states at $k_z = 0$ and $k_z = \pi$ are shown in (c) and (d), respectively.}
\label{fig:surf}
\end{figure}

We explicitly verify the above physical picture in bulk $\gamma$-graphyne. To do this, we construct an \emph{ab-initio} Wannier tight-binding model from the first-principles band structure. The surface spectra are computed based on this model. In Fig.~\ref{fig:surf}(a), we plot the spectrum for the surface normal to $\hat{y}$. One can see that there exist some surface bands insides the bulk gap, as indicated by the arrows. However, unlike conventional topological insulators, they do not cross the band gap, so that the surface spectrum is gapped. This is consistent with our expectation above.

Next, we investigate the existence of hinge modes. Here, we take a tube-like geometry as shown in Fig.~\ref{fig:surf}(c, d). The structure is periodic along $z$ and a cross section contains $30\times 30$ conventional cells (32,400 sites in total). The obtained spectrum plotted in Fig.~\ref{fig:surf}(b) is a band structure in $k_z$. One observes an almost flat band at Fermi level ($E=0$). We checked that this band is actually doubly degenerate and its states are localized on the hinges. In Fig.~\ref{fig:surf}(c) [Fig.~\ref{fig:surf}(d)], we plot the spatial distribution of the two degenerate states at $k_z=0$ ($k_z= \pi$), which confirms that they are located at the two hinges connected by $\mathcal{PT}$.

As we mentioned, the hinge band for 3D RCI can be viewed as formed by stacking the corner modes of 2D RCI layers along $z$. In the previous work Ref.~\cite{chen2022}, it was found that for 3D graphdiyne, the interlayer coupling is relatively strong, such that the bulk gap closes (and the bands are inverted) at certain $k_z=\pm k_c$. The resulting system is a real nodal-line semimetal and its topological hinge modes only appears in the interval of $(-k_c,k_c)$~\cite{chen2022}. In contrast, for 3D $\gamma$-graphyne, the interlayer coupling is sufficiently weak such that a global band gap is maintained. Therefore, it becomes a 3D RCI and the topological hinge modes exist across the whole 1D BZ of $k_z$.


\section{Topological phase transition}

We have established that bulk $\gamma$-graphyne is a 3D RCI, formed by stacking  graphyne monolayers (which are 2D RCIs~\cite{chen2021}) with relatively weak interlayer coupling. Then, a natural thought would be: If one enhances the interlayer coupling, there might appear a topological phase transition.

A direct method to enhance the interlayer coupling is to decrease the interlayer distance by applying a lattice strain.
Here, by using the conventional cell in Fig.~\ref{fig:structure}(e), we vary the interlayer distance $d$ and study its effects on the band structure.

\begin{figure}[tb]
  \includegraphics[width=8 cm]{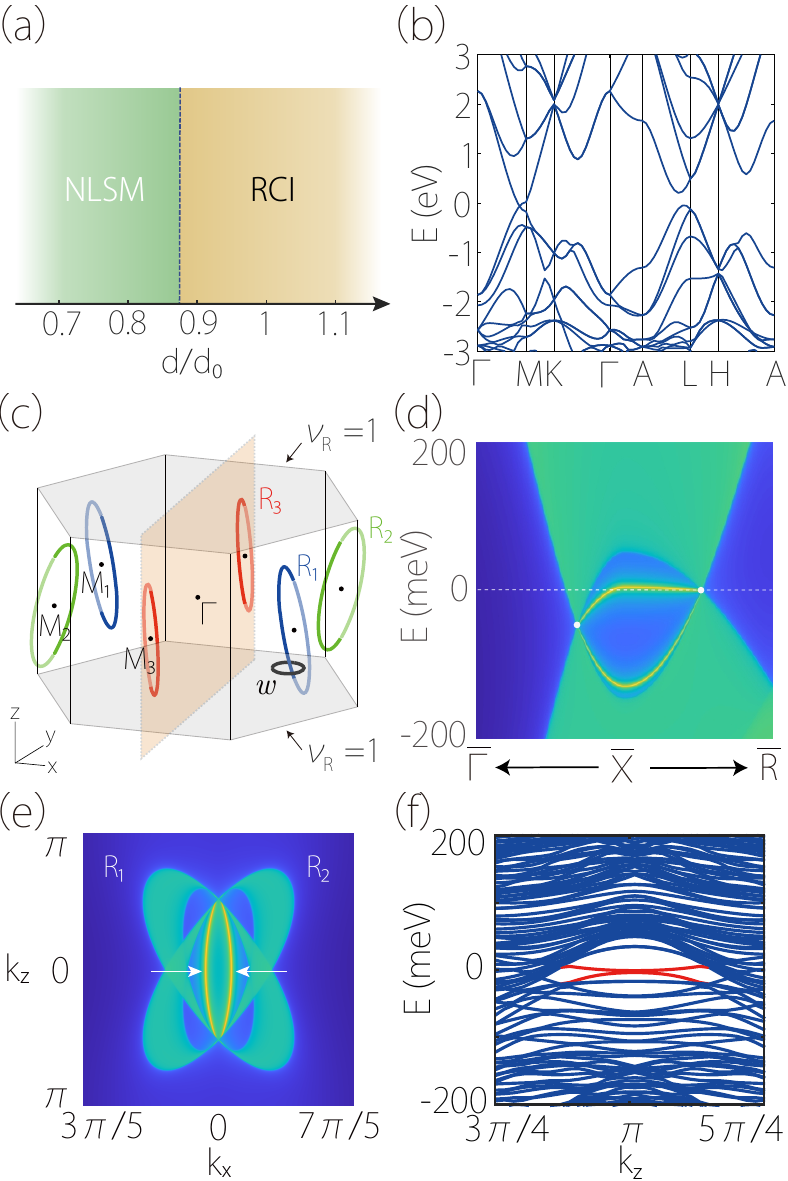}
  \caption{(a) Phase diagram of bulk $\gamma$-graphyne when varying the interlayer distance. NLSM stands for the nodal-line semimetal phase. (b) Band structure when the interlayer distance is decreased to $0.85 d_0$. Here, the plot is with respect to the conventional cell. (c) Illustration of the three nodal rings for the state in (b). The rings are centered around the three $M$ points. Each ring lies in a vertical mirror plane and features a nontrivial 1D invariant $w$. The $k_z=\pi$ plane is still a 2D RCI with $\nu_R=1$. (d) Projected spectrum for the side surface normal to the $y$ direction (see Fig.~1(f)). White dots correspond to points on the projected nodal ring. (e) shows the constant energy slice at Fermi level for (d). The arrows indicate the drumhead type surface states. (f) Spectrum for the nodal-line semimetal state with the tube-like geometry in Fig.~\ref{fig:surf}(c). The red color indicates the hinge modes.  }
\label{fig:stress}
\end{figure}

The obtained phase diagram is shown in Fig.~\ref{fig:stress}(a). Here, $d_0$ denotes the experimentally determined equilibrium interlayer distance.
The result shows that a transition indeed occurs when the interlayer spacing is decreased to about $\sim 0.87 d_0$. After the transition, the system becomes gapless and is a nodal-line semimetal. It should be pointed out that
a compressive strain $>10\%$ on a 3D material could be difficult to realize in practice. Nevertheless, since the physics
discussed below are interesting and should be common for similar topological systems, we shall proceed to study the phase after transition, despite the practical difficulty to achieve it in $\gamma$-graphyne.

In Fig.~\ref{fig:stress}(b), we plot the band structure for $d=0.85d_0$. The system is semimetallic, and one observes a band crossing on the $\Gamma$-$M$ paths. A careful scan over the BZ shows that band crossing points actually form three nodal rings centered at the three $M$ points on the boundary of BZ. ({We have checked that the number of rings is also three if plotted in the BZ for the primitive cell.}) Each ring is located in a vertical mirror plane and is protected by the mirror symmetry. In addition, there is another protection for the nodal ring from the quantized Berry phase under $\mathcal{PT}$. This gives a 1D topological charge
\begin{equation}
  w=\frac{1}{\pi} \oint_{C} \operatorname{Tr} \mathcal{A}(\boldsymbol{k}) \cdot d \boldsymbol{k} \quad \bmod 2
\end{equation}
for each nodal ring, where $C$ is a small loop encircling the ring and $\mathcal{A}$ is the Berry connection for the occupied bands. It should be noted that the nodal rings here are quite different from those found in 3D graphdiyne. In
3D graphdiyne, there are a pair of nodal rings connected by $\mathcal{T}$ (or $\mathcal{P}$), and besides the 1D charge $w$, each ring also has a nontrivial 2D topological charge $\nu_R=1$ defined on a sphere $S^2$ enclosing the ring~\cite{chen2022}. In comparison, there are three rings in Fig.~\ref{fig:stress}(c), and each ring is its own $\mathcal{T}$ (or $\mathcal{P}$) partner. Importantly, here, each ring must have a trivial 2D topological charge. Otherwise, a 2D (fixed $k_z$) slice above the rings and a 2D slice below the rings would have a different $\nu_R$, which is a contradiction as the BZ is continuous [see Fig.~\ref{fig:stress}(c)].

Next, we consider the boundary spectra of this real nodal-ring semimetal state. In Fig.~\ref{fig:stress}(d),
we plot the surface spectrum for the side surface normal to $\hat{y}$. One can clearly observe the drumhead type surface bands which are connected to points on the nodal ring (the white dots).
In Fig.~\ref{fig:stress}(e), we plot the constant energy slice at Fermi energy for this side surface BZ.
The two bulk nodal rings $R_1$ and $R_2$ are projected onto this surface and form the petal shaped pattern in Fig.~\ref{fig:stress}(e). The remaining nodal ring $R_3$ is in a mirror plane normal to this surface, so it gets projected to a straight line at $k_x=0$ (not shown in this figure). Inside the projected nodal rings $R_1$ and $R_2$, one observes the drumhead type surface states as indicated by the arrows.

As for the topological hinge modes, we expect they would still exist in the $k_z$ interval around $k_z=\pi$. This is because the 2D $k_z$-slices there are still gapped and maintain a nontrivial $\nu_R=1$. In Fig.~\ref{fig:stress}(f), we plot the spectrum for the same tube-like sample geometry in Fig.~\ref{fig:surf}(c), which indeed confirms our expectation. It should be pointed out that the two hinge bands begin to split when moving away from $k_z=\pi$. This is due to the finite-size effect: The local gap for these $k_z$-slices decreases with the deviation from $k_z=\pi$; this decreased gap increases the spatial spread of the hinge modes and hence the coupling of the two hinges for a finite-size sample, leading to enhanced splitting.

\section{model for 3D graphyne}

\begin{figure}[tb]
  \includegraphics[width=8cm]{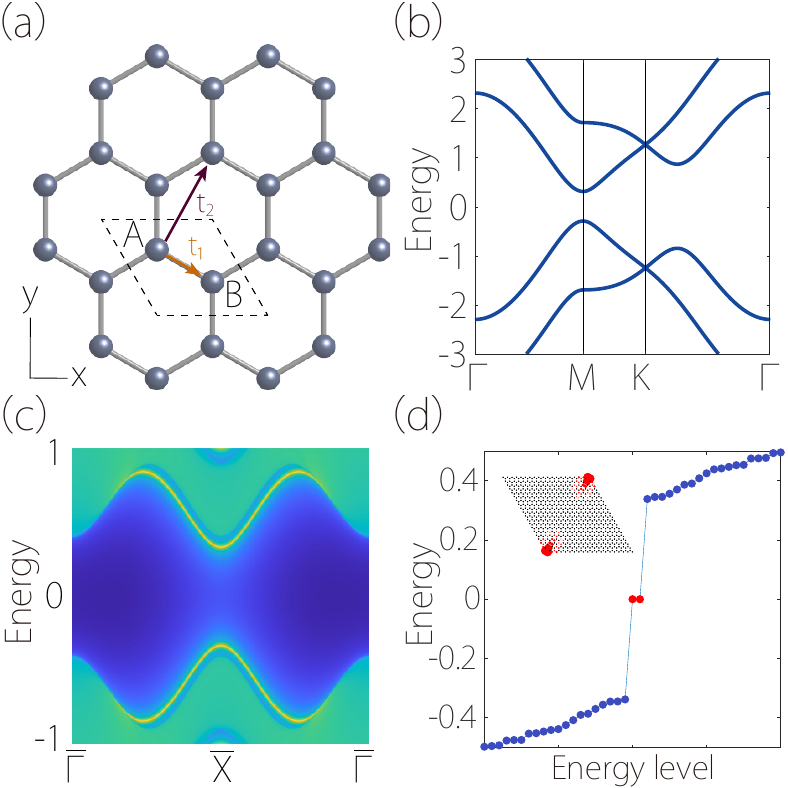}
  \caption{(a) Illustration of our minimal model for a single $\gamma$-graphyne layer. It is a honeycomb lattice with two orbitals per site. We consider hoppings up to second neighbor.  (b) A typical band structure for our model Eq.~(\ref{H2d}). (c) Projected spectrum of the model for an edge normal to $\hat{y}$. (d) Spectrum of the model with a disk geometry in the inset. The inset also shows the distribution of the two zero-modes. In the calculation, we set the parameters as: $t_1 = 1$, $\varepsilon = 0.7$, and $t_{2} = 0.2$.}
  \label{fig:TB2D_Fourband}
  \end{figure}

To better understand the topological features and to facilitate future studies,
we develop a simple effective model for bulk $\gamma$-graphyne.

Since the system is a vdW layered material, we adopt a layer construction approach by starting with a monolayer model. 
In Ref.~\cite{sheng2019}, we presented a $k\cdot p$ effective model for monolayer graphdiyne, expanded at $\Gamma$ point. However, for monolayer graphyne, its low-energy bands involve three valleys at the three $M$ points~\cite{chen2021}. To capture this structure, it is more proper to build a tight-binding model. To conform with the symmetry of the system, we consider a honeycomb lattice shown in Fig.~\ref{fig:TB2D_Fourband}(a). Each unit cell has two sites $A$ and $B$, and we put two different orbitals at each site. The symmetry of monolayer $\gamma$-graphyne has $D_{3d}$ point group symmetry. We take the following
representations for the generators of this group:
\begin{equation}
  \mathcal{P} = \sigma_z\tau_x,\qquad {C}_{2y} = \sigma_z\tau_0,\qquad C_{3z} = \sigma_0\tau_0.
\end{equation}
where the Pauli matrices $\sigma$ and $\tau$ represent the orbital and the sublattice degrees of freedom, respectively. The time-reversal symmetry is represented as $\mathcal{T} = \mathcal{K}$, with $\mathcal{K}$ the complex conjugation. Under the symmetry constraint, we construct the following tight-binding model using the approach of Ref.~\cite{zhang2022b}:
%
%
\begin{equation}\label{H2d}
  \begin{aligned}
  \mathcal{H}_\text{2D} = & \varepsilon\sigma _{z}\tau _{0}+t_1(\sin{\frac{\sqrt{3}k_{y}}{3}}-2\cos{\frac{k_{x}}{2}}\sin{\frac{\sqrt{3}k_{y}}{6}})\sigma _{z}\tau _{y}\\
  &+t_1(\cos{\frac{\sqrt{3}k_{y}}{3}}+2\cos{\frac{k_{x}}{2}}\sin{\frac{\sqrt{3}k_{y}}{6}})\sigma _{z}\tau _{x}\\
  &+2t_2(\sin{k_{x}}-2\sin{\frac{k_{x}}{2}} \cos{\frac{\sqrt{3}k_{y}}{2}})\sigma_{y}\tau _{0},\\
  \end{aligned}
  \end{equation}
%
where $\varepsilon$ represents an energy splitting between the two orbitals, $t_1$ describes the nearest neighbor hopping which is diagonal in the orbital basis, $t_2$ is for the second neighbor hopping processes (see Fig.~\ref{fig:TB2D_Fourband}(a)), and all model parameters are real valued. In Fig.~\ref{fig:TB2D_Fourband}(b), we plot a typical band structure for this model, which shows a semiconductor with direct band gap at the three $M$ points, capturing the essential features in a single $\gamma$-graphyne layer. More importantly, we find that with $t_1 = 1$, $|\varepsilon|<1$, and a small but nonzero $t_2$, the model describes a 2D RCI with topological corner zero-modes, as verified in Fig.~\ref{fig:TB2D_Fourband}(c, d).


\begin{figure}[htb]
  \includegraphics[width=8cm]{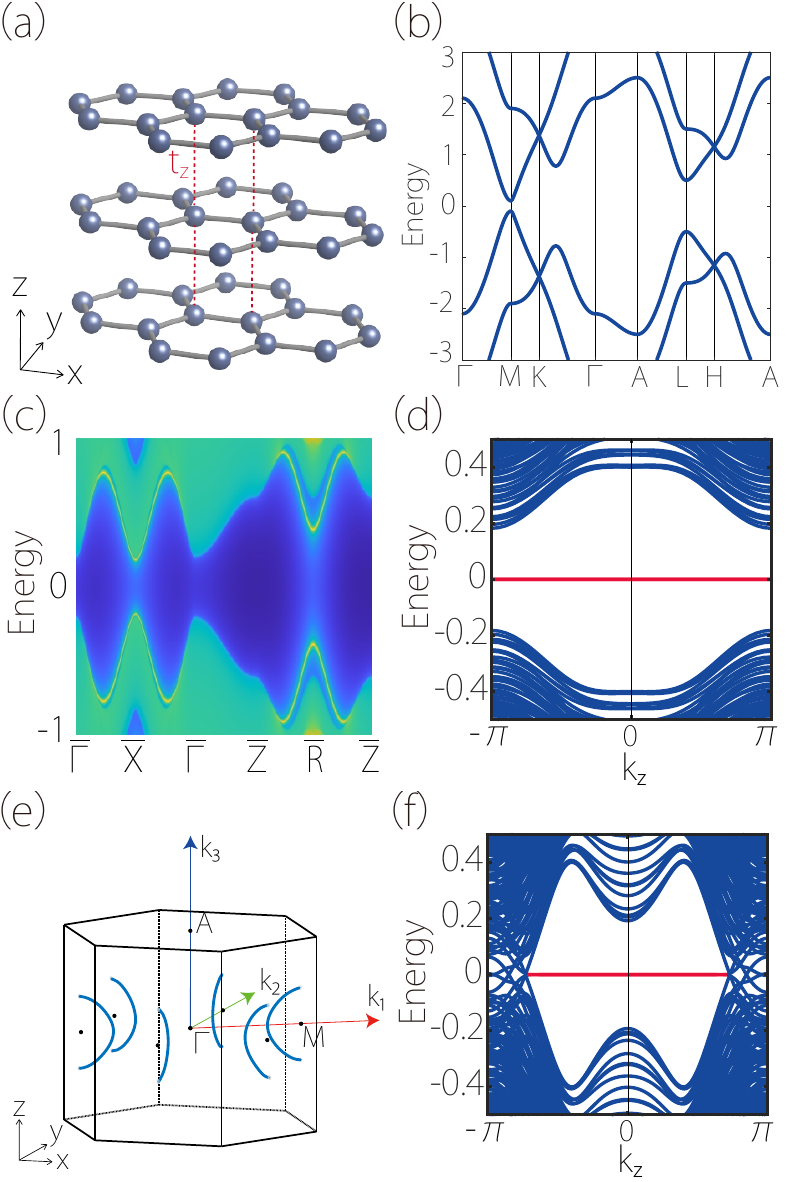}
  \caption{(a) Illustration of our 3D model, which is obtained by simply stacking the 2D models along $z$. (b) Typical band structure for our model Eq.~(\ref{H3d}) in the 3D RCI phase. (c) Projected spectrum for the side surface norm to $\hat{y}$. (d) Spectrum for the tube-like geometry (similar to Fig.~3(c)). The red color indicates the hinge modes. (e) By increasing the interlayer coupling, the system can transition into the nodal-line semimetal phase. The figure illustrates the three nodal rings formed in the model. (f) shows the corresponding spectrum for the tube-like geometry.  Here, we set the parameters as:  $t_1 = 1$, $\varepsilon = 0.7$, $t_{2} = 0.2$, $t_z = 0.1$ for (b-d) and $t_z = 0.3$ for (e, f).}
  \label{fig:TB3D_Fourband}
  \end{figure}

Then, we stack this 2D model into a 3D tight-binding model as in Fig.~\ref{fig:TB3D_Fourband}(a). Constrained by the $D_{3d}$ symmetry, we add a simple interlayer coupling term to reach the following 3D model:
%
%
\begin{equation}\label{H3d}
  \begin{aligned}
    \mathcal{H}_\text{3D}=  \mathcal{H}_\text{2D}+
    2t_z\cos{k_z}\sigma_z\tau_0,
  \end{aligned}
\end{equation}
where $t_z$ is the interlayer hopping strength.

We expect that if we start from the 2D RCI model, with weak $t_z$, the corresponding 3D model would remain a semiconductor and form a 3D RCI. This is confirmed by our calculation result in Fig.~\ref{fig:TB2D_Fourband}(b). Particularly, the obtained surface spectrum in Fig.~\ref{fig:TB3D_Fourband}(c) and the hinge zero-modes in Fig.~\ref{fig:TB3D_Fourband}(d) capture the essential features of bulk $\gamma$-graphyne in Fig.~\ref{fig:surf}(a, b).
Now, if we increase the value of $t_z$, there indeed emerges a phase transition into a nodal-ring semimetal state. In this state, there are three nodal rings centered at the three $M$ points, as illustrated in Fig.~\ref{fig:TB3D_Fourband}(e), reproducing the features in Fig.~\ref{fig:stress}(c). The topological hinge modes in this case exist in the $k_z$ interval around $k_z=\pi$, as shown in Fig.~\ref{fig:TB3D_Fourband}(f), which also agrees with the \emph{ab initio} result in Fig.~\ref{fig:stress}(d).

The above discussion demonstrates that our proposed minimal model successfully captures the rich topological features in bulk $\gamma$-graphyne. This can serve as a good starting point for the studies on 3D RCI.

\section{Discussion and conclusion}

High-quality bulk $\gamma$-graphyne has already been realized in experiment. The sample size can reach a few microns~\cite{hu2022}.
Hence, our predictions here can be directly tested in experiment. The bulk band structure and the surface spectrum can be imaged by using the angle resolved photoemission spectroscopy (ARPES)~\cite{lv2021}. The topological hinge modes can be detected by
the scanning tunneling spectroscopy (STS), as previously demonstrated in the study of Bi~\cite{zheng2018,schindler2018}.

In studying the topological hinge modes, we took the sample geometry in Fig.~\ref{fig:surf}(c, d). It should be mentioned that the existence of such topological zero-modes is general. As demonstrated in previous studies, their existence is not sensitive to the geometry nor any small perturbations to the system. Particularly, for the 3D RCI here, as long as the sample maintains $\mathcal{PT}$ symmetry, there must be topological zero-modes existing at certain pairs of hinges connected by $\mathcal{PT}$.

In conclusion, we have revealed the recently synthesized bulk $\gamma$-graphyne as a realistic example of 3D RCIs.
We show that bulk $\gamma$-graphyne can be regarded as formed by 2D RCIs stacked along the $z$ direction. Its bulk is characterized by the real Chern number $\nu_R=1$ for each $k_z$-slice of BZ. The bulk real topology manifests as topological hinge zero-modes, which exist on the whole 1D BZ. Lattice strain that enhance the interlayer coupling can drive a phase transition into a real nodal-line semimetal phase. We show that distinct from 3D graphdiyne, the nodal rings here
do not occur in pairs and they have a trivial 2D topological charge. Finally, we construct a minimal model,
which captures all above mentioned topological features in this system. Our work discovers the nontrivial topology in bulk $\gamma$-graphyne and promotes it as a promising platform for exploring fascinating physics of real topological phases.


\begin{acknowledgments}
We thank D. L. Deng for helpful discussions. We acknowledge the computational support from HPC of the Beihang University and Texas Advanced Computing Center. This work is supported by National Key R\&D Program of China (2022YFA1402600), the NSFC (Grants No. 12174018, No. 12074024), the Fundamental Research Funds for the Central Universities, and the Singapore Ministry of Education AcRF Tier 2 (T2EP50220-0026).

\end{acknowledgments}

\appendix
 \counterwithin{figure}{section}
 \counterwithin{table}{section}
 \counterwithin{equation}{section}

\section{Computation Method}\label{appendix:method}
Our first-principles calculations were carried out based on the density-functional theory (DFT), as implemented in the Vienna \textit{ab initio} simulation package (VASP)~\cite{kresse1994,kresse1996}. The ionic potentials were treated by using the projector augmented wave method~\cite{blochl1994}. The band structure results presented in the main text are based on the HSE06 approach~\cite{krukau2006}. Other approaches to the exchange-correlation functional, such as Perdew-Burke-Ernzerhof (PBE)~\cite{perdew1996}, modified Becke-Johnson potential (mBJ)~\cite{rasanen2010}, and HSE+GW~\cite{onida2002}, were also tested (see Supplemental Material). The qualitative results, especially the topological characters, are unaffected by the different choices.
The plane-wave cutoff energy was set to 500 eV.
The Monkhorst-Pack $k$-point mesh~\cite{monkhorst1976} of size $8\times8\times 8$ was used for the BZ sampling. The surface spectra were calculated by first constructing an \emph{ab initio} tight-binding model based on the maximally localized Wannier functions~\cite{marzari1997,souza2001} and then by using the surface Green's function method~\cite{lopezsancho1984,lopezsancho1985} implemented in the WannierTools package~\cite{wu2018a}.
The  \emph{ab initio} tight-binding model was also used in the calculation for the tube-like sample geometry.

\bibliographystyle{apsrev4-2}
\bibliography{3dGPY}


\end{document}